\documentstyle[11pt]{article}

\newcommand{\z}{\zeta}\newcommand{\la}{\lambda}
\newcommand{\de}{\delta}\newcommand{\ve}{\varepsilon}
\newcommand{\e}{\epsilon}
\begin{document}
\vspace{0.5cm}
\centerline{\Large\bf Weinberg propagator}
\centerline{\Large\bf of a free massive particle with an arbitrary spin}
\centerline{\Large\bf from the BFV--BRST path integral}
\vspace{1cm}
\centerline{\large{\bf V.G.Zima}$^1$ and {\bf S.O.Fedoruk}$^2$ }
\begin{center}{$^1$ {\it Kharkov State University,\\
 310077 Kharkov, 4 Svoboda Sq., Ukraine\\
   e-mail: zima@postmaster.co.uk}\\
$^2$ {\it Ukrainian Engineering--Pedagogical Academy,\\
 310003 Kharkov, 16 Universitetskaya Str., Ukraine\\
e-mail: fed@postmaster.co.uk}}
\end{center}
\vspace{1.2cm}
\begin{abstract}
 The transition amplitude is obtained for a free massive particle of arbitrary
spin by calculating the path integral in the index--spinor formulation within the
BFV--BRST approach. None renormalizations of the path integral measure were
applied. The calculation has given the Weinberg propagator written in
the index--free form with the use of index spinor. The choice of boundary
conditions on the index spinor determines holomorphic or antiholomorphic
representation for the canonical description of particle/antiparticle spin.
\end{abstract}

PACS: 04.60.Gw, 03.70.+k

{\it Keywords:} Spinning particle; BFV--BRST path integral; Weinberg
propagator

\newpage
\section{Introduction}
In the development of extended object theories, the problem of covariant
description of spinning particles, in particular, the problem of covariant
quantization of these particles, plays a double role. On the one hand, it is
an educational model which allows one to illustrate the progress achieved and
to train oneself in application of the developing methods. On the other hand,
it is a starting point and, in a certain sense, the desired result of these
theories intended to realize consistently the fundamental quantum and
relativistic principles, so that one could speak about the construction of the
interaction theory for particles remaining the only actually observable
manifestation of the fundamental structure of matter.

An important part of the quantization problem of the particle
is the calculation of its
propagator. The most powerful modern method for solving this problem,
just as for the problem of quantization in general, is the BFV-BRST
approach~\cite{1}.  Howerever, up to now calculations of transition amplitudes
for massive spinning particles in this approach have been carried out only
in rather limited number of papers~\cite{2} in the framework of
pseudoclassical mechanics~\cite{BerMar,Casal} and were
restricted to the spin $1/2$.
For earlier paper with the old method of the calculation see,
for example~\cite{HennT}.
For field strength of massless particle there
are calculations of the propagator for
arbitrary spins~\cite{Papad} and also in  pseudoclassical formulation.

In this paper we apply the BFV-BRST quantization procedure to the free
arbitrary-spin massive particle moving in the space-time of the
dimension $D=4$. As in above-cited
papers the question is in an obtaining
equivalent of known expressions by novel methods and finding for an
adequate representation of the result. In our opinion, the approach
accepted here to the
description of spin in terms of the index spinor~\cite{3}
is very helpful in solution of the quantization problem. In view of an
universal character of such description and relative novelty of the index
spinor conception we give some details of our construction.

It is well known that at the classical level the spin is putting in the
theory by introducing a few of additional coordinates, a part of which can
be auxiliary. These spin variables can be commuting or
anticommuting Lorentz scalars of
the target space-time, spinors, vectors ets. Among them one can
extract subsets describing
the internal geometry of the particle world line
and spinors of the internal
symmetry group, if such is present. If the postulated group of
space-time symmetry is wider than the ordinary one, the several Lorentz
representations can be collected into more complicated formations such as
(super)twistors of (super)conformal group. A choice of using spin
variables is in essence the matter of convenience.
It may be useful to have various formulations of free
spinning particle mechanics and then to study how interaction
are ``switched on'' in each case.

Nevertheless, the spinors as basic spin variables are especially
attractive. Actually, in the space of states the notion of spin is
associated with the space
of irreducible representation of the small group.
In the case of a massive particle
it is natural to take the rest frame momentum
as a standard one. Then the small group is the group
of the space rotations ${\rm SO}(3)$ or its quantum mechanical counterpart,
i.  e.  the spinor group ${\rm Spin}(3)\approx {\rm SU}(2)$. Topologically
irreducible representations of this group, as for each other compact
group, are finite dimensional and since the group is linear
they have the canonical realization in the space of multilinear forms (or
tensorial degrees of the fundamental representation).
An irreducible representation is
determined by its highest weight or, equally, a Young tableau or
by an eigenvalue of the Casimir operator, which is namely the spin for the
case. The Young tableau visually determines
the degree and the
type of symmetry for the multilinear form, in particularly, the number of
its different spinor arguments, which is equal to the rank
of the small group $r=1$.
This circumstance points to the preferred and
fundamental role of the set of variables which consists of $r$ spinors
of small group in the ``mechanical'' spin theory.

In the relativistic
theory such a set of variables retains its role because the
quantum mechanical group of
rotations ${\rm SU}(2)$ has, as the proper complex envelope, the
quantum mechanical Lorentz group ${\rm SL}(2,{\bf C})$,
which is thus the relevant hipercompact complex group. So the space
of a unitary representation of the rotational group is simultaneously the
space of a nonunitary finite-dimensional representation of the Lorentz
group.  Thus an application of the Weyl  ``unitary trick'' connects
irreducible representations of the compact group ${\rm SU}(2)$ with
analytic and antianalitic irreducible representations
of the corresponding complex
group ${\rm SL}(2,{\bf C})$.

Until recently, the bosonic spinors have been used
mostly as twistor--like variables that resolve
the mass constraint in the massless case~\cite{Penr}. However, the
potential of these variables is far from being exhausted. For example,
in the theory with bosonic spinors there is, at least at the classical level,
a very simple solution of the problem of infinite reducibility of
fermionic $\kappa$--symmetry due to a possibility of construction
of the projectors with such spinors.
Its connection with the solution in the
framework of doubly supersymmetric models~\cite{SorTV}
is still unclear. This
circumstance justifies a further analysis of bosonic--spinor particle
models since it implies a possible existence of more subtle geometric
and group--theoretical aspects.

In construction of the mechanical theory of particle spin the well known
Borel-Weyl-Bott theorem is in a certain sense the Ariadnian thread . The
theorem states that for any given irreducible representation
of any compact connected Lie
group there exists a classical dynamical system which quantization yields
this representation as the quantum Hilbert space. In spin theory that
group is the spinor group ${\rm Spin}(3)$. The factual
construction of the mechanical system mentioned
in the theorem for the massive
particle with spin can be realized in the rest frame in two ways by using
of either commuting (bosonic) coordinates or anticommuting (fermionic or
Grassmann) ones.
We have as bosonic
construction the first order Lagrangian of the form
$\alpha i(\zeta\dot{\bar\zeta}-\dot\zeta\bar\zeta )
-\lambda (\zeta\bar\zeta -1)$,
where a dot denotes derivative with respect to the development parameter
$\tau$, $\lambda $ is a Lagrange multiplier, $\alpha $ is
a real constant and
standard index free notations are used to contraction of the dimensionless
spinor $\zeta$ and its complex conjugate $\bar\zeta$. In the kinetic
term the minus sign provides its irreducibility to a total derivative. The
potential term, i. e. the spin constraint $\zeta\bar\zeta -1 \approx 0$
entering the action,
restricts the configuration space to
the group manifold because then the $2\times 2$ matrix with the lines
$\zeta$ and $\epsilon\bar\zeta$ is a unimodular unitary one. Here
$\epsilon$ is the unit antisymmetric spinor.

The pair of the complex conjugated primary constraints
$p_\zeta \approx -i\alpha\bar\zeta$ and
$p_{\bar\zeta} \approx i\alpha\zeta$ belongs to the second class. Upon
quantization with the use of Dirac brackets, $\zeta$ and $\bar\zeta$ are
realized up to a multiplier as bosonic creation-annihilation operators
$\{\bar\zeta ,\zeta\}_{DB} =-i/2\alpha$. The spin constraint
$\zeta\bar\zeta -1 \approx 0$ fixes
the value of the ``particle number operator''
up to an ordering constant. Upon quantization the eigenvalue $J$ of the
modulus of the angular momentum $M_i =\alpha\zeta\sigma_i \bar\zeta$ must be
half-integer. Thus, under quantization bosonic theory leads to the states
of arbitrary spin because the ordering constant is indetermined.

The fermionic Lagrangian can be taken quite similar
$\alpha i(\theta\dot{\bar\theta}-\dot\theta\bar\theta )$
without any potential term. Here $\theta$ denotes an odd spinor. This
results in spins up to $1/2$ over a spinless ground state.
To extract a definite spin it should be introduced the spin constraint
$\theta\bar\theta =0$ also but now there is only
a finite set of eigenvalues of
the  corresponding quantum operator.

Using of bosonic and fermionic spinor variables
simultaneously one can take the spinless vacuum,
then the spin of bosonic subsystem is regarded as the spin of the Clifford
vacuum for the fermionic subsystem.

To obtain the relativistic extension of these models one
should construct a Lagrangian whose ``spinor part'' in the rest frame is reduced
to the expressions discussed above. This is achieved by an obvious
transformation of the kinetic and potential parts
$$
\alpha i(\zeta\dot{\bar\zeta}-\dot\zeta\bar\zeta )\to
i(\zeta\hat p\dot{\bar\zeta}-\dot\zeta\hat p\bar\zeta )\, ,\quad
(\zeta\bar\zeta -1)\to (\zeta\hat p\bar\zeta -j)\, .
$$
Here $p$ is the energy-momentum vector,
$\hat p$ is its contraction with the Pauli
matrices $\sigma$ so that in the rest frame
$\hat p =m\sigma_0$, where $m$ is the particle mass.
In these conversions some natural redefinitions have been
made and the spinor acquires the dimension and becomes a Weyl spinor.
We call this spinor the index spinor because of the role which it plays after
quantization.  The particle Lagrangian arises after adding the kinetic
$p\dot x$ term and the potential $-\frac{e}{2}(p^2 +m^2 )$ one of the
free spinless particle for coordinates of the phase space.
Here $e$ is an einbein. In this way one
obtains the action of the paper~\cite{3}, where the sign of the
particle energy coincides with the sign of the  ``classical spin'' $j$
due to the spin constraint $\zeta\hat p\bar\zeta -j =0$.  The spectrum of
the model consists of the induced
representations of the Poincar{\' e} group.

The kinetic term of this model
can be written in terms of the bosonic superform
$\omega =dx-i(d\zeta\sigma\bar\zeta -\zeta\sigma d\bar\zeta )$
and obviously possesses
bosonic supersymmetry $\zeta\to\zeta +\varepsilon$,
$x\to x+i(\zeta\sigma\bar\varepsilon -\varepsilon\sigma\bar\zeta )$, where
$\varepsilon$ is a  constant commuting spinor.  This symmetry is destroyed
by the spinorial potential term in the action.

With fermionic coordinates Casalbuoni-Brink-Schwarz supersymmetric
action~\cite{Casal,7}
appears, if the constraint for extracting a particular value of spin has not
been inserted.
In known way~\cite{AzLuk,GIKOS} this model can
be generalized to the model with extended supersymmetry by introducing of
isospinor coordinates and attaching indices of internal group to the odd
spinor coordinates. The terms correspond to the central charges also can
be added to the action.

In both, bosonic and fermionic, cases
the massless particle appears when on takes the consistent
limit $m\to 0$.
Certainly, all these models are theories with the first and
second class constraints. But in the presence of bosonic spinor
coordinates (which are not nilpotent) there is no problem with
the covariant and irreducible separation of constraints
into first and second classes, at least
at the classical level, because we can construct the projectors by using
spinors $\zeta$ and $\hat p\bar\zeta$ for which
$\zeta\hat p\bar\zeta =j$.

There are other
approaches~\cite{Ferb}-\cite{KuzLS}
which use commuting spinors as variables
for the description of spin. First of all it is the use of entries of the
fundamental representation matrix. Often these variables are called
harmonics by some misuse of language. Literally the harmonics~\cite{GIKOS}
are pure auxiliary gauge variables which parametrize an arbitrary
frame with respect to the canonical one and can be also
regarded as a bridge connecting the
representations of the group with ones of its subgroup.
They acquire dynamical status only if a gauge is fixed and some basic
variables transmit to them a part of their functions.
So the use of these
variables as dynamical ones should be understood as such a choice of
gauge in which some initial dynamical variables have been gauged away and
their role passed to the harmonics. In principle
together with the first class constraints,
which provide the harmonics with a gauge nature,
their matrix should be subject to constraints which place it
in the corresponding group  (in higher space-time dimensions it is
impossible to formulate these constraints
as conditions of the conservation of quadratic forms).
Thus the theory with harmonics is strongly restricted.  Careful
account of these constraints in the quantization procedure is often rather
nontrivial~\cite{ZimF}.  Therefore, sometimes
the consideration is carried out in
the frame of a quasiharmonic approach with
dynamical ``harmonics''~\cite{Band}.
One takes
some number of independent harmonic spinors assuming that implicit
gauge conditions and second class constraints have been resolved with
respect to other ones.
Undoubtedly, any set of independent spinors for the construction of
arbitrary irreducible representation
can be found among lines or rows of the matrix of spinor
representation because any representation can be realized in the space of
functions on the group. From the index spinor point of view, in such
a consideration there is an implicit use of all or a part of index
spinors.  The index variables form
a system of independent quantities in terms of which one can construct
the harmonic matrix taking into account all present restrictions.
Nonclassical nature of the spinor group in higher
dimensions~\cite{Nissim,Galp} is a powerful
evidence on behalf of explicit exploitation
of the index variable conception
not the complete and consecutive harmonic one.

In the massless case adapting the harmonic frame of reference to the
positive energy-momentum vector, i. e. directing one of its basic vector
along the isotropic vector of energy-momentum, it is possible to resolve
the mass constraint $p^2\approx 0$
in terms of the harmonic spinors $v$ and $\bar v$ as
$p\sim v\sigma\bar v$. Then,
after suitable gauge fixing a dynamical role of space-time variables is
given to harmonics
$v$ and their canonical conjugate momenta.
In lower critical dimensions of space-time one has
succeeded to deal with spinors subjected to explicitly formulated
constraints.  Twistor formulations~\cite{Penr,Ferb,EisSol,Eisenb,Berk}
arise just in this way
(here we are interested in the connection of twistors with harmonics
and not in their
group theoretical aspects).  Of course, introducing twistors, when it
is possible in the stated sense, one should not follow
the described scheme
necessary, i. e. one can take twistors irrespectively to harmonics. In
particular, it is possible to introduce the twistors in parallel
to the index spinors which then can be gauged away.
Only in this case one can obtain the classical twistor theory
in a unique way. In
the theory without index spinor, where the sign of energy is
indeterminated, a choice of sign is necessary in such
a transition.

It is important that even the use of pure gauge harmonics
essentially changes
the situation, since it yields a topologically
nontrivial configuration space in the case of pure gauge torus degrees of
freedom.  Precisely this makes it possible to obtain different spins in the
massless case without introducing nongauge variables~\cite{ZimF}.

Mechanical systems, which describe a massless particle with arbitrary
spin, have the same number of dynamical degrees of freedom as the system
for spinless particle. Therefore at the classical level all the models with
commuting spinors as basic spin variables can be sufficiently easy reduced
to each other by fixing some gauge symmetries. But it would be
incorrect to think that all these models are identical. Only in the approach
with the index spinor the massive and massless particles of arbitrary spin
have uniform description with natural generalization to higher space-time
dimensions.

For example, if in the massless index spinor
model~\cite{3} one identified the
variables $v= |j|^{-1/2}p_\zeta$ and $\omega = |j|^{+1/2}\zeta$
as spinorial components of the twistors~\cite{EisSol}  then the
first class spin constraint
$S\equiv \frac{i}{2} (\zeta p_\zeta -\bar p_\zeta \bar\zeta) -j \approx 0$
has been rewritten as the twistorial Hamiltonian
$H\equiv \frac{i}{2} (\omega v -\bar v \bar\omega )-j \approx 0$
with a ``classical helicity'' $j$.
The fundamental twistor constraints
$T_{\alpha\dot\alpha}\equiv p_{\alpha\dot\alpha}-
v_\alpha \bar v_{\dot\alpha} =0$,
which solve the massless condition $p^2 =0$,
can be projected onto the twistor spinors.
The projections
$vT\bar v$, $vT\bar \omega$, $\omega T\bar v$ and $\omega T\bar \omega$
are equivalent to the set of constraints
of the theory with the index spinor~\cite{3}
which consists the massless constraint $p^2\approx 0$ and
a part of projections of the spinorial constraints
$d_\zeta\equiv ip_\zeta -\hat p\bar\zeta\approx 0$
and $\bar d_\zeta\equiv -ip_{\bar\zeta} -\zeta\hat p\approx 0$
onto the index spinors
$\zeta$ and $\hat p\bar\zeta$, i. e. to the constraints
$p^2$, $\zeta \hat p\bar d_\zeta $,
$d_\zeta \hat p\bar\zeta $ and
$\frac{1}{2} (\zeta d_\zeta +\bar d_\zeta \bar\zeta) \approx
-(\zeta\hat p\bar\zeta -j)$.
The projection
$\frac{1}{2} (\zeta d_\zeta -\bar d_\zeta \bar\zeta)
=\frac{i}{2} (\zeta p_\zeta +\bar p_\zeta \bar\zeta) $
which falls out from the last listing is nothing
but the conformal constraint
$\omega v +\bar v \bar\omega =0$ for which the constraint
$\omega T\bar \omega =0$
plays a role of gauge condition and vice versa.

It is intersecting that in terms of twistorial variables
$v$ and $\omega$ the spinor constraints
$d_\zeta$ and $\bar d_\zeta$ take the form
$v\sim \hat p\bar\omega$ and c. c. which is in a sense dual to the twistor
conditions $\omega =\hat x\bar v$ and c. c.
We can say that the index spinors $\zeta$, $p_\zeta$ and the twistor ones
$\omega$, $v$ replace each other under  Fourier transformation
in massless particle description.

The index spinor can be added to pseudoclassical
mechanics~\cite{BerMar,Casal}. Then
on the mass shell this theory becomes classically equivalent to the theory
describing spin by both commuting and anticommuting spinors
simultaneously. By now such a theory have not
been developed enough. So we would
like only to point out some of its interesting peculiarities and a way of
establishing the equivalence. For simplicity we restrict ourself to the case
of pseudoclassical mechanics with the single anticommuting vector
$\psi_\mu$ usually reffered to as describing spin $1/2$ particle.
It is useful to represent the anticommuting variables of
the pseudoclassical mechanics in the form
$$
\psi_{\alpha\dot\alpha}\equiv
\psi_{\mu}\sigma^\mu_{\alpha\dot\alpha}=
2j^{-1/2}[(\hat p\bar\zeta )_\alpha
\bar\theta^\prime_{\dot\alpha} + (\zeta \hat p)_{\dot\alpha}
\theta^\prime_{\alpha} ] +2\rho\zeta_\alpha\bar\zeta_{\dot\alpha}\, ,
$$
$$
\psi_5 =-j^{-1/2}\frac{p^2}{m}
[\zeta\theta^\prime + \bar\theta^\prime\bar \zeta ] +
\frac{1}{m}\rho(\zeta\hat p\bar\zeta) +\tilde\psi_5
\, ,
$$
where primed thetas $\theta^\prime$, $\bar\theta^\prime$
are anticommuting, as well as $\rho$ and $\tilde\psi_5$.
These representations for
$\psi_{\alpha\dot\alpha}$, $\psi_5$ are general on the surface
of the constraints
$\zeta\hat p\bar\zeta \approx j \ne 0$ and
$p^2 \approx -m^2 \ne 0$.
The quantities $\rho$ and $\tilde\psi_5$ are unabiguously defined by
$\psi_\mu$ and $\psi_5$, respectively.
The spinor $\theta^\prime_\alpha$ is defined up to a term of the form
$i(\hat p\bar\zeta)_\alpha \psi$ with the anticommuting $\psi$.
Since now we have $p^\mu \psi_\mu =m\tilde\psi_5$,
the main constraint
$p^\mu \psi_\mu +m\psi_5 \approx =0$ of the pseudoclassical mechanics
takes an easy solved form
$\tilde\psi_5 \approx 0$.
Furser we suppose $\tilde\psi_5 =0$.

Under the substitution of these expressions into the kinetic term
$\frac{i}{2}(\psi^\mu \dot\psi_\mu +\psi_5 \dot\psi_5 )$
for the anticommuting (pseudo)vector $\psi_\mu$ and (pseudo)scalar $\psi_5$
we use also  for the index spinor,
$2i\dot\zeta\hat p +\lambda\zeta\hat p =0$ and its c. c.,
the equations of motion for the
energy-momentum vector, $\dot p =0$, and the trivial identities
$\zeta^2 =\bar\zeta^2 =0$.
One should redefine $\theta^\prime$ and $\zeta$ by the mutually
conjugated phase multipliers
$k$ and $\bar k =k^{-1}$ in order to obtain the new index spinor
$\zeta^\prime =k\zeta$, which satisfies to the equations
$\dot\zeta^\prime\hat p = 0$, and the new anticommuting spinor
$\theta =k^{-1}\theta^\prime$, which satisfies to the equations
$\dot\theta = k^{-1}(\dot\theta^\prime -\frac{\lambda}{2i}
\theta^\prime )$. The equation for $k$ is
$\dot k =-\frac{\lambda}{2i}k$, and it can be easily solved as
$k=C\exp (-\frac{1}{2i}\int^{\tau}_{\tau_0}\lambda d\tau)$,
where $C$ is a constant of integration, $\tau_0$ is an initial
moment of the ``time'' $\tau$, and $\lambda$ is a
Lagrange multiplier at the spin constraint
$\zeta\hat p\bar\zeta -j$.

After these redefinitions one obtains  for the kinetic term
an expression in which is contained the new anticommuting spinors
$\theta$, $\bar\theta$ and the anticommuting scalar $\rho$ with
their derivatives. The constant spinors
$\zeta^\prime$, $\bar\zeta^\prime$ enter in this expression as well.
It is instructive to note that
$$
\dot\psi_{\alpha\dot\alpha}=
2j^{-1/2}[(\hat p\bar\zeta^\prime )_\alpha
\dot{\bar\theta}_{\dot\alpha} + (\zeta^\prime \hat p)_{\dot\alpha}
\dot\theta_{\alpha} ] +
2\dot\rho\zeta^\prime_\alpha\bar\zeta^\prime_{\dot\alpha}\, , \quad
\dot\psi_5 =j^{-1/2}m
[\zeta^\prime\dot\theta + \dot{\bar\theta}\bar \zeta^\prime ] +
\frac{1}{m}\dot\rho j
\, .
$$

Regarding the expression as a Lagrangian one can find the equation of
motion for
$\rho$ which is $j^{3/2}\dot\rho =m^2 (\zeta^\prime\dot\theta +
\dot{\bar\theta}\bar\zeta^\prime )$.
This equation can be easily integrated but it is not required to the
substitution of its solution into the action. The direct use of the
equation of motion for $\rho$ in the Lagrangian yields the expression
$$
i(\theta\hat p\dot{\bar\theta} -\dot\theta\hat p\bar\theta) +
\frac{im^2}{j}(\dot\theta\zeta^\prime \, \zeta^\prime\theta +
\dot{\bar\theta}\bar\zeta^\prime \, \bar\zeta^\prime\bar\theta )\, .
$$

Here
the first term originates from the vector part  of the initial
kinetic term of the pseudoclassical mechanics
only and is nothing but the
spinor kinetic term of the CBS superparticle~\cite{Casal,7}.
The second term originates from the
both items in the kinetic term of the pseudoclassics and represents
a term which corresponds to the second-rank-spinor central charge of
superparticle. Tensor central charges in particle models
have been considered in~\cite{BanLuS}.  In our case we have a complex
self-dual antisymmetric isotropic (singular) tensor of second rank.

For the massless particle the pseudoclassical description contains only
anticommuting vector
$\psi_\mu$. Here we have
$p^\mu \psi_\mu =-j\rho$. So the imposition of the constraint
$p^\mu \psi_\mu \approx 0$ yields
$\rho\approx 0$ and the fermionic $\kappa$-symmetry of the model is
achieved without involving any central charge.
The further calculations in the massless case a quite similar to those
have been made for the massive particle.
Because  of the gauge
equivalence of the massless particle model,
with the index spinor and with the twistor, which was
mentioned above, our calculation can be regarded as analogous to
those in the paper~\cite{VolZ} but without a direct appeal to the
twistors.

In this paper we obtain the propagator of the
free arbitrary-spin massive $D=4$ particle as the BFV-BRST path integral.
The present scheme for the description of spin
in terms of the index spinor~\cite{3}, used in this paper, is obviously
applicable to both the massless case and the case of higher space--time
dimensions; so the problem we deal with is only
a test to estimate the
efficiency of the approach.

The consideration of this type within the framework of modern
quantization methods is performed for the first time.
Along with the extension
to higher spins, the advantages of the Hamiltonian formulation
have been first
used for such a problem to full extent and the path
integral has been calculated
without resorting to arbitrary uncontrolled renormalizations of
the integration measure. The derived propagator coincides with that found
previously within the traditional field theory in the framework of
($2J+1$)--component formalism~\cite{4}.

We make no recourse to the conversion of second--class constraints, because
it would be natural to carry out this consideration when studying the massless
case, where the bosonic $\kappa$--symmetry of the model leads to a nontrivial
algebra of first--class constraints.

The choice of the domain of integration over the gauge degrees of freedom,
being the key point in a similar consideration, is made by finding and
choosing the fundamental region of the modular group in the Teichm\"{u}ller
space~\cite{20}. This choice is not associated with the ambiguity
of the procedure, it is rather the selection of a solution of the problem
out of the set of possible ones for a fixed system. As a result, the
causal propagator arises naturally.

A careful analysis of boundary conditions requires
the modification of the
expression for the transition amplitude
in the path integral form by adding the
boundary terms to the classical action~\cite{5}. The presence of
second--class constraints gives rise to the canonical conjugation between the
index spinor and its complex--conjugate one.
Therefore, the boundary conditions are different
for them, i. e., one is fixed at the initial moment of time, and the other is
fixed at the final moment. It is shown that the resulting alternative
corresponds to the choice of the particle spin description: either holomorphic with
undotted spinors or antiholomorphic with dotted ones. The transition from one
choice to an other is equivalent to the exchange
of the roles between particles
and antiparticles.

This article is organized as follows. In sect.~2 we discuss the classical
formulation of a spinning particle with the index spinor, proposed for the first
time in paper~\cite{3}, and carry out the Hamiltonian analysis in
such a framework, which is necessary for the quantum path--integral
consideration. In sect.~3 we construct the BFV--BRST path--integral expression
for the transition amplitude in the ``relativistic'' gauge and evaluate it
in sect.~4 for the holomorphic and antiholomorphic boundary conditions
on the index--spinor variables. This calculation includes the determination
of the integration domain and properly the integration over Teichm{\" u}ller
parameters. In sect.~5 we establish the links between the obtained
transition amplitude and the propagator of a massive arbitrary--spin
particle in the $(2J+1)$--component formalism of the conventional field theory.

Here we use the spinor conventions of ref.~\cite{6}.

\section{Classical consideration of a spinning particle with the index spinor}
In the usual space--time ($D=4$), a spinning particle can be described with the
commuting coordinates $(z^A)=(x^\mu,\z^\alpha,\bar\z^{\dot\alpha})$, where
$x$ is a four--vector and $\z$ is a Weil index spinor. We write the
Lagrangian of the particle in the form~\cite{3}
\begin{equation}\label{1}
L=p\dot\omega-\frac{e}{2}(p^2+m^2)-\la(\z\hat p\bar\z-j)\,,
\end{equation}
where the bosonic ``superform'' is
$$
\omega\equiv\dot\omega\,d\tau=dx-id\z\sigma\bar\z+i\z\sigma d\bar\z\,.
$$
The kinetic term $p\dot\omega$ represents the sum of the standard kinetic term
for the spinless particle $p\dot x$, where $p_\mu$ is an auxiliary energy--momentum
vector, and the spinning addend, which takes the standard oscillator form
$im(\z\sigma_0\dot{\bar\z}-\dot\z\sigma_0\bar\z)$
in the rest frame. As a result, the form $\omega$ coincides with
the N=1 SUSY superform, if one replaces the Grassmannian spinor by the
index one there. It should be stressed that this coincidence is not the
result of some direct or naive generalization of the well--known expressions
of the supersymmetric theory. Actually, this circumstance reflects an
essential common feature of spin descriptions in terms of commuting
and anticommuting variables. Namely, both these descriptions arise quite
directly as neat ``relativizations'' of well known representations of the
small group in terms of $c$--numbers and $a$--numbers, respectively, i. e.,
by construction of the corresponding induced representations of the
Poincar{\' e} group. In natural way this inducing leads to the bosonic and
fermionic supersymmetry of the respective kinetic terms in the language
of theoretical mechanics. ``Unfortunately'' the bosonic supersymmetry is
destroyed by the necessary restriction of the bosonic configuration space
imposed by the spin constraint~\cite{3}; the ``relativistic'' form
$\z\hat p\bar\z-j\approx 0$ of this constraint is explicitly involved in
the Lagrangian~(\ref{1}) with the Lagrange multiplier.

The einbein $e$ and $\la$ are the Lagrange multipliers in the
Lagrangian~(\ref{1}). The dimensionless constant $j$ is the classical spin
the sign of which determines the sign of energy. Our action
$$
A=\int_{\tau_i}^{\tau_f}L\,d\tau
$$
universally describes  both massless and massive cases, but in this work
we restrict ourselves to consideration of the massive particle only, so that
$m^2>0$. In the absence of the last term in the Lagrangian~(\ref{1}), the
massless particle action coincides with the Casalbuoni\---Brink\---Schwarz
action~\cite{Casal,7} if one will interpret $\z$ as the Grassmannian
spinor.

Apart from the constraints inserted into the action explicitly, i. e.,
the mass constraint
\begin{equation}\label{2}
T\equiv\frac{1}{2}(p^2+m^2)\approx 0
\end{equation}
and the spin one
\begin{equation}\label{3}
\z\hat p\bar\z-j\approx 0\, ,
\end{equation}
the Hamiltonization~\cite{8} of the theory reveals the spinor
Bose--constraints as well
\begin{equation}\label{4}
d_\z\equiv ip_\z-\hat p\bar\z\approx 0\, ,\qquad
\bar d_\z\equiv -i\bar p_\z-\z\hat p\approx 0\, .
\end{equation}
We omit obvious first class constraints on the momenta, which are
canonically conjugate to the Lagrange multipliers, and the second--class
constraints on the momenta conjugated to auxiliary variables $p$.
Accounting of the last constraints in the strong sense by introducing the
Dirac brackets is trivial and does not modify the brackets for fundamental
variables. On the constraints surface the spin constraint~(\ref{3}) is
equivalent to the following
\begin{equation}\label{5}
S\equiv S_\z-j\equiv\frac{i}{2}(\z p_\z-\bar p_\z\bar\z)-j\approx 0\, ,
\end{equation}
since $S\equiv\frac{1}{2}(\z d_\z-\bar d_\z\bar\z)+(\z\hat p\bar\z-j)$.

The fundamental brackets are  $\{z^A,p_B\}=\de^A{}_B$;
$\bar p_\z\equiv p_{\bar\z}$.

The constraint algebra is found immediately, its nontrivial brackets are
\begin{equation}\label{6}
\{ d_\z,\bar d_\z\}=2i\hat p\, ,\qquad \{ S,d_\z\}=\frac{i}{2}d_\z\, ,
\qquad \{S,\bar d_\z\}=-\frac{i}{2}\bar d_\z\, .
\end{equation}
Thus, the constraints $(F_a)=(F_1,F_2)\equiv (T,S)$ belong to the first class,
whereas the spinor constraints $(G_i)=(d_{\z\alpha},\bar d_{\z\dot\alpha})$
are the second--class ones. The latter implies the consideration of the
nonzero mass particle, i. e., $\hat p\tilde p=m^2>0$. Certainly in the procedure of
Hamiltonization, the spinor constraints~(\ref{4}) are primary, whereas the mass
constraint~(\ref{2}) and the spin one~(\ref{3}) are the constraints of the second
step of the procedure.
The total Hamiltonian is a linear combination of the first--class constraints.
This is due to the reparametrization invariance of the action.

The mass constraint~(\ref{2}) generates usual reparametrizations of space--time
coordinates in the phase space
$$
\de x^\mu =p^\mu\e,\qquad \de p_\mu =0,\qquad \de e=\dot\e\, ,
$$
where the last equality follows from the invariance condition of the Hamiltonian
action up to surface terms.

The spin constraint~(\ref{5}) generates phase transformations of phase space
coordinates (in a sense of multiplying by the phase multiplier)
$$
\de\z^\alpha=\frac{i}{2}\z^\alpha\varphi,\qquad
\de p_{\z\alpha}=-\frac{i}{2}p_{\z\alpha}\varphi\, ;\qquad
\de\la=\dot\varphi\, .
$$

The corresponding variation of the action
$$
\de A=\frac{1}{2}(p^2-m^2)\e\Bigl|_{\tau_i}^{\tau_f}+
 j\varphi\Bigr|_{\tau_i}^{\tau_f}
$$
vanishes solely if $\e(\tau_i)=\e(\tau_f)=0$ and $\varphi(\tau_i)=
\varphi(\tau_f)$. This circumstance makes directly admissible only
``relativistic gauges''~\cite{1}, i. e., the gauges with derivatives which
impose restrictions on $\dot e$, expressing it in terms of other phase
space variables. Then the condition of gauge conservation leads to the
second--order equation on the parameter $\e$, which has the unique solution for any
appropriate boundary conditions~\cite{9}. To use a canonical gauge
without derivatives, one should
consider it as a singular limit of a succession of admissible gauges~\cite{10}
or introduce appropriate boundary terms in the Hamiltonian action~\cite{5}.

\section{BFV--BRST path integral for the transition amplitude}
The most profound method for calculation of transition amplitude for
constrained systems is the BFV--BRST formalism~\cite{1}. In this approach,
for each first--class constraint $F_a$ the set of coordinates of
the initial phase space is
supplemented by ``dynamical'' Lagrange multipliers

$(\la^a)\equiv(\la_T,\la_S)$ with the same Grassmannian parity, their
canonically conjugate momenta $\pi_a$, $\{\la^a,\pi_b\}=\de^a_b$, and
the ghost variables of the opposite parity. The ghost sector contains
Grassmannian odd ghosts $C^a$, antighosts $\tilde C_a$ and their canonically
conjugate quantities $\tilde {\cal P}_a$ and ${\cal P}^a$,
$\{ C^a,\tilde {\cal P}_b\}=\{ {\cal P}^a,\tilde C_b\}=\de^a_b$. The variables
$\la$, $\pi$, $C$, ${\cal P}$ are real, whereas $\tilde {\cal P}$, $\tilde C$
are pure imaginary.

The variables of original phase space are subjected to the second--class
constraints~(\ref{4}), but the algebra of the first--class constraints $F_a$
remains Abelian even after introducing the Dirac brackets
$$
\{A,B\}_D =\{ A,B\}-\frac{i}{2p^2}\{ A,\bar d_\z\}\tilde p\{ d_\z,B\}+
(-1)^{AB}\frac{i}{2p^2}\{ B,\bar d_\z\}\tilde p\{ d_\z,A\}\, .
$$
Thus, the BRST charge has a zero rank and is a linear combination of the first--class
constraints, $F_a$ and $\pi_a$, of the extended phase space
\begin{equation}\label{7}
\Omega=F_a C^a  +\pi_a {\cal P}^a \, ;
\end{equation}
$$
\{\Omega,\Omega\}=\{\Omega,\Omega\}_D=0 \, ,\qquad
\overline{\Omega}=\Omega\, .
$$
The BRST charge is Grassmannian odd, $\e(\Omega)=1$, and has the ghost number
one, ${\rm gh}(\Omega)=1$, as it is supposed that
$$
{\rm gh}(C)={\rm gh}({\cal P})=-{\rm gh}(\tilde{\cal P})=-{\rm gh}(\tilde C)=1 \, .
$$

The path integral for the transition amplitude
\begin{equation}\label{8}
Z_\Psi=\int D[z,p_z;\la,\pi;C,\tilde {\cal P};{\cal P},\tilde C]
\prod_{i,\tau}\de(G_i)\prod_\tau (2\pi)^2\left|\det\{G_i,G_j\}\right|^{1/2}
\exp (iA_{eff})\, ,
\end{equation}
includes the usual Liouville measure. Let us describe it in more detail.
This means that in the standard finite--dimensional
approximations of the path integral, the product of differentials
of each pair of the canonically conjugate real
bosonic variables in the measure is divided by $2\pi$. The differential of
each variable that remains without its pair, in accordance with the boundary
conditions under consideration, is also divided by $2\pi$. Here, all
what has been said relates to the
variables $p_\mu$ and $\la^a$. Similar multipliers are absent for
the Grassmannian quantities. In the Hamiltonian approach, the multipliers
corresponding to the realification Jacobian of the using complex variables
do not appear in the measure.

Fulfillment of the second--class constraints~(\ref{4})
in expression~(\ref{8}) is provided
by the functional $\de$~--~functions. The multipliers corresponding to the
realification Jacobian do not arise in the product $\prod\limits_i\de(G_i)$ of
$\de$~--~functions of the complex second--class constraints. The measure is
normalized by the determinant of Poisson brackets matrix for the second--class
constraints
$$
\det\{ G_i,G_j\}=\left(\det\{ d_\z,\bar d_\z\}\right)^2=
\left( 4p^2\right)^2\, .
$$
In addition, for every time ``moment'' $\tau$ the factor $2\pi$ should be
introduced into the measure on each pair of real bosonic second--class constraints.

The effective Hamiltonian action is
\begin{equation}\label{9}
A_{eff}=\int\limits_{\tau_i}^{\tau_f}\left( p\dot x+\dot\z p_\z+
\bar p_\z\dot{\bar\z}+\pi\dot\la+\tilde{\cal P}\dot C+\tilde C\dot {\cal P}-
H_\Psi\right)d\tau+A_{b.t.}\, .
\end{equation}
The choice of the BRST Hamiltonian $H_\Psi$ and the boundary term $A_{b.t.}$
is argued below.

For the theory with a reparametrization invariance, the BRST Hamiltonian $H_\Psi$ is
the BRST ``derivative'' of the gauge fermion $\Psi$:
$$
H_\Psi=\{\Omega,\Psi\}\, .
$$
In the amplitude~(\ref{8}), one may use on equal footing both Poisson
and Dirac brackets because, in our case, the Poisson brackets of the first--class
constraints (entering into $\Omega$) and the arbitrary function are different
from the Dirac brackets by addends which are proportional to the second--class
constraints only. Thus these terms vanish on the second--class constraint
surface. The gauge fermion
is Grassmannian odd, $\e(\Psi)=1$, pure imaginary, $\overline\Psi=-\Psi$,
and has a negative ghost number, ${\rm gh}(\Psi)=-1$. As it is known~\cite{1}, the transition
amplitude does not depend on the choice of a gauge fermion if the path integral
is taken over the paths which belong to the one  class of equivalence
with respect to the BRST transformation.
Such class is extracted by choosing the appropriate gauge and
boundary conditions. The relativistic gauge with derivatives for the Lagrange
multipliers ($\dot\la^a=0$) corresponds to
\begin{equation}\label{10}
\Psi=\tilde {\cal P}_a\la^a\, ,
\end{equation}
then
\begin{equation}\label{11}
H_\Psi=F_a\la^a+\tilde {\cal P}_a {\cal P}^a \, .
\end{equation}
It should be stressed that an attempt to simplify further the expression for
$\Psi$ by excluding some addends is rather undesirable.
In such a way one loses the restriction to the only
equivalence class of the paths and, as a result, arrives at
``averaging'' over many classes. Then an infinite
renormalization of the integration measure becomes necessary~\cite{11}.

We carry out the calculation of transition amplitude in the coordinate representation
for the variables $z^A$ and in the mixed representation for the ghosts, i. e.,
we choose the boundary conditions
\begin{equation}\label{12}
x^\mu(\tau_i)=x_i{}^\mu\, ,\quad x^\mu(\tau_f)=x_f{}^\mu \, ;
\end{equation}
\begin{equation}\label{13}
\z^\alpha(\tau_1)=\z_1{}^\alpha\, ,\quad
\bar\z^{\dot\alpha}(\tau_2)=\bar\z_2{}^{\dot\alpha}\, ,
\end{equation}
where the marks $(1,2)$ of spinors must be understood as $(f,i)$ for the
holomorphic choice and as $(i,f)$ for the antiholomorphic one;
\begin{equation}\label{14}
\pi_a(\tau_i)=\pi_a(\tau_f)=0;\quad C^a(\tau_i)=C^a(\tau_f)=0;\quad
\tilde C_a(\tau_i)=\tilde C_a(\tau_f)=0.
\end{equation}
The boundary values are not fixed for the rest of variables. The boundary
conditions imposed are BRST--invariant and ensure vanishing of the
BRST charge on the boundaries. This provides the form--invariance of
amplitude~(\ref{8}). One can understand the vanishing of the boundary values of
the BRST charge as a classical manifestation of the quantum condition
$\hat\Omega\left| \psi_{phys}\right>=0$~\cite{12}.  The choice
of boundary conditions
for the index spinor is covariant and consistent with
canonical conjugacy of $\z$ and $\bar\z$ (appearing due to the
second--class constraints). Such choice is not unique. Using combinations
of the index spinor and its conjugate momentum with other variables of the phase
space, one can propose a variety of covariant boundary conditions on index
variables. All they are  in essence equivalent  and reflect
a concrete choice of the quantum description of a spin (i. e., realization
of the Hilbert space of quantum states). We restrict ourselves to the
consideration of two basic variants~(\ref{13}).  As the simplest ones,
they are described in the literature now~\cite{3}.

With boundary conditions~(\ref{13}), the correctness of the variational
principle, i. e., independence of any variation of the action from the
boundary values of the variation for variables which are not fixed at the
boundary, needs introducing the boundary term
\begin{equation}\label{15}
A_{b.t.}=-\frac{\varepsilon_{\z}}{2}(\z_i p_{\z i} +\z_f p_{\z f}
-\bar p_{\z i}\bar\z_i -\bar p_{\z f}\bar\z_f)\, .
\end{equation}
Here $\varepsilon_\z=+1$ corresponds to the holomorphic choice of the boundary
condition~(\ref{13}) and $\varepsilon_\z=-1$ corresponds to the antiholomorphic
one.

\section{Calculation of the path integral}
In the gauge~(\ref{10}), the path integral~(\ref{8}) is factorized
\begin{equation}\label{16}
Z_\Psi = Z\cdot Z_{gh}\, .
\end{equation}
The path integral over the odd ghost variables has a simple Gaussian form
\begin{equation}\label{17}
Z_{gh}=\int D[C,\tilde {\cal P};{\cal P},\tilde C]\cdot
\exp\{i\int\limits_{\tau_i}^{\tau_f}
(\tilde {\cal P}_a\dot C^a -\dot{\tilde C}_a {\cal P}^a
-\tilde {\cal P}_a{\cal P}^a)d\tau\} \, .
\end{equation}
In Eq.~(\ref{17}), integration by parts has been performed in the index of
the exponent with the boundary conditions for $\tilde C$ being taken into
account.  This integral can be calculated by partition of the variation interval
for the evolution parameter $\tau$ into $N$ equal parts. Put $T_\tau
\equiv\tau_f -\tau_i$ and $\Delta\tau=T_\tau/N$.  Now the integrations over
${\cal P}$ and $\tilde {\cal P}$ automatically determine the normalization
multiplier $(i\Delta\tau)^{2N}$ for the measure in the intermediate integral
\begin{equation}\label{18}
Z_{gh} =\int \tilde D[C, \tilde C]\exp\{-i\int_{\tau_i}^{\tau_f}{}
\dot{\tilde C}_a\dot {C}^a{}d\tau\}
\end{equation}
in its calculation by a discretization of the interval
$[\tau_i,\tau_f]$. One can directly obtain~(\ref{18}) from~(\ref{17})
without discretization by sequential integrations over $\tilde {\cal P}$,
which creates the $\de$~--~function $\de({\cal P}-\dot C)$, and over ${\cal
P}$, which annihilates this $\de$--function, if no care is taken for
normalization. When the vanishing boundary values of the ghost variables
$C$ and $\tilde C$ are not assumed, the result of integration in~(\ref{18})
has the form
\begin{equation}\label{19}
Z_{gh}=-T_\tau^2\exp\{-i(\tilde
C_{fa}-\tilde C_{ia})(C_f^a-C_i^a)/T_\tau\}\, .
\end{equation}

Let us give some details of integration over the ghosts. As the integrand
in~(\ref{17}) does not include any cross terms with the ghosts for different
constraints, it is sufficient to restrict the consideration by
the case of a unique first--class constraint. We have
$$
Z_{gh}^{(1)}=\lim_{N\to\infty}Z_{gh}^{(1)}[C_i,C_f;\tilde C_i,\tilde C_f;T_\tau,N]
$$
where
\begin{eqnarray}
Z_{gh}^{(1)}[C_i,C_f;\tilde C_i,\tilde C_f;T_\tau,N]&=&
\int{}\prod_{k=1}^{N-1}dC_k d\tilde C_k\cdot\prod_{k=1}^{N}
d\tilde {\cal P}_k d{\cal P}_k\cdot\nonumber\\
&&\cdot\exp\left\{i\sum_{k=1}^{N}\left[\tilde {\cal P}_k(C_k-C_{k-1})-
(\tilde C_k-\tilde C_{k-1}){\cal P}_k-
\tilde {\cal P}_k {\cal P}_k \Delta\tau \right]\right\} \, ;\nonumber
\end{eqnarray}
$C_0=C_i$, $\tilde C_0=\tilde C_i$, $C_N=C_f$, $\tilde C_N=\tilde C_f$. The
superscript in $Z_{gh}^{(1)}$ refers to the case of a unique constraint.

The shifts $\tilde{\cal P}_k\to\tilde{\cal P}_k-(\tilde C_k-\tilde C_{k-1})/
\Delta\tau$, ${\cal P}_k\to{\cal P}_k+(C_k-C_{k-1})/\Delta\tau$ make possible
integrations over ${\cal P}_k$ and $\tilde{\cal P}_k$:
\begin{eqnarray}
Z_{gh}^{(1)}[C_i,C_f;\tilde C_i,\tilde C_f;T_\tau,N]
&=&i\Delta\tau\int\prod_{k=1}^{N-1}dC_k d\tilde C_k(i\Delta\tau)\cdot\nonumber\\
&&\cdot\exp\left\{-\frac{i}{\Delta\tau}\sum_{k=1}^{N}(\tilde C_k-\tilde C_{k-1})
(C_k-C_{k-1})\right\}\, .\nonumber
\end{eqnarray}

As it is easily varified by the induction in $N$,  this integral is independent
of $N$:
$$
Z_{gh}^{(1)}[C_i,C_f;\tilde C_i,\tilde C_f;T_\tau,N]=
iT_\tau\exp\{-i(\tilde C_f-\tilde C_i)(C_f-C_i)/T_\tau\}=Z_{gh}^{(1)}.
$$

For zero boundary values of $C$ and $\tilde C$~(\ref{13}), and even for
weaker conditions $C_f=C_i$ or $\tilde C_f=\tilde C_i$, we have
\begin{equation}\label{20}
Z_{gh}=-T_\tau{}^2\, .
\end{equation}

Thus the transition amplitude is
\begin{eqnarray}
Z_\Psi&=&-T_\tau{}^2\int D[z,p_z;\la,\pi]\prod_{i,\tau}
\de(G_i)\cdot\prod_\tau 4|p^2|(2\pi)^2\cdot \nonumber\\
&&\cdot\exp\left\{i\int\limits_{\tau_i}^{\tau_f}
\left( p\dot x+\dot\z p_\z+\bar p_\z\dot{\bar\z}+\pi\dot\la
-F_a\la^a\right) d\tau+iA_{b.t.}\right\}\, ,\label{21}
\end{eqnarray}
where only the path integration over even variables remains to be done.

The integrals over the momenta $\pi_a$ of the Lagrange multipliers $\la^a$ give
the $\delta$~--~functions $\delta(\dot \la^a)$. So, after the path integration over
$\la^a$ is performed, only usual integrals over zero modes of $\la^a$
remain in $Z_\Psi$. A precise determination
of integration domain over zero modes of Lagrange multipliers, which plays a key
role in our consideration, will be considered below.

It is convenient to carry out the integration by parts
in the index of the exponent in~(\ref{21}):
$\int\limits_{\tau_i}^{\tau_f}p\dot xd\tau=px\biggl|_{\tau_i}^{\tau_f}-
\int\limits_{\tau_i}^{\tau_f}\dot pxd\tau$. Then the path integrals over $x$
give the $\delta$~--~functions $\delta(\dot p)$,
so that the path integrals over
$p$ are reduced to usual integrals over zero modes of $p$.
Hence, instead of the
considered integrals in the index of the exponent,  the expression
$ip(x_f-x_i)$ appears.

The second--class constraints~(\ref{4}) have the form solved with respect to the
spinor momenta $p_\z$ and $\bar p_\z$. So, we can easily integrate over these
variables, using the functional $\delta$~--~functions in the measure.

Now the transition amplitude~(\ref{21}) takes the form
\begin{equation}\label{22}
Z_\Psi=-T_\tau{}^2\int\frac{d^4 p}{(2\pi)^4}e^{ip(x_f-x_i)}
\frac{d\la_T d\la_S}{(2\pi)^2}\exp\left\{-i\frac{T_\tau}{2}\la_T(p^2+m^2)
+iT_\tau\la_S J\right\}\cdot Z_\z
\end{equation}
with the path integrations over the index spinor
\begin{equation}\label{23}
Z_\z=\int\prod_\tau d^2 \z d^2 \bar\z |p^2|\cdot\exp\left\{
i\int\limits_{\tau_i}^{\tau_f}
\left(-i\dot\z\hat p\bar\z +i\z\hat p\dot{\bar\z}-
\la_S \z\hat p\bar\z\right)d\tau +
i\tilde A_{b.t.} \right\}
\end{equation}
being factored. The boundary term~(\ref{15}) acquires the form
\begin{equation}\label{24}
\tilde A_{b.t.}=-i\ve_\z (\z_i\hat p\bar\z_i +\z_f\hat p\bar\z_f )\, .
\end{equation}
The quantum spin $J$ is introduced in~(\ref{22})
(or rather from the very beginning in~(\ref{8})) instead of $j$
to stress the possibility
of redefinition of the classical value of spin $j$ by an ordering
constant in the construction of a quantum theory corresponding to the classical
one~(\ref{1}). As in the original functional Liouville
measure~(\ref{8}), in Eq.~(\ref{22}) the differential of each
coordinate of zero mode of the energy--momentum vector $p$ is divided by
$2\pi$. The same concerns the differential of zero mode of
each Lagrange multiplier.

As usually, the exponential multiplier in the expression for Gaussian
integral~(\ref{23}) can be  easily found by the saddle--point method.
When the boundary conditions are taken into account, the extremality
of the exponent index with
respect to $\bar\z$ and $\z$ is reached on the equations of motion
for $\z$ and $\bar\z$:
\begin{equation}\label{25}
2i\dot\z\hat p +\la_S \z\hat p =0\qquad\mbox{and}\qquad\mbox{c.\,c.}
\end{equation}
Only the boundary term contributes to the integrand exponent~(\ref{23})
after Eqs.~(\ref{25}) are taken into account.
With boundary conditions~(\ref{13})
 the solutions of equations~(\ref{25}) take the form
\begin{equation}\label{26}
\z\hat p=e^{\frac{i}{2}\la_S(\tau-\tau_1)}\z_1\hat p
,\qquad\hat p\bar\z =e^{-\frac{i}{2}\la_S(\tau-\tau_2)}\hat p\bar\z_2 \, .
\end{equation}
Thus the integral~(\ref{23}) acquires the form
\begin{equation}\label{27}
Z_\z=\exp\{2\ve_\z \z_1\hat p\bar\z_2 e^{-i\ve_\z\la_S T_\tau /2}\}\, .
\end{equation}

The pre--exponential multiplier in~(\ref{27}) can be found from the
prelimiting expression in the equation
$Z_\z=\lim\limits_{N\to\infty}Z_\z[\z_1,\bar\z_2;T_\tau,N]$
for calculation of the considered Gaussian path
integral~(\ref{23}) by discretization of the interval for the development
parameter $\tau$.
For example, in the holomorphic case, we have
\begin{eqnarray}
Z_\z[\z_f,\bar\z_i;T_\tau,N]&=&\int\prod_{k=1}^N d^2\z_{k-1}d^2\bar\z_k
|p^2|/\pi^2\cdot \nonumber \\
&&\cdot\exp\left\{-2\left(1+i\la_S\frac{T_\tau}{2N}\right)
\sum_{k=1}^N\z_{k-1}\hat p\bar\z_k+
2\sum_{k=0}^N\z_k\hat p\bar\z_k\right\} \nonumber
\end{eqnarray}
with $\bar\z_0=\bar\z_i$, $\z_N=\z_f$.
Using mathematical induction it is not difficult to verify that
$$
Z_\z[\z_f,\bar\z_i;T_\tau,N]=\left[1+(\frac{\la_S T_\tau}{2N})^2\right]^{-N}
\exp\left\{2\z_f\hat p\bar\z_i\left(1+i\la_S\frac{T_\tau}{2N}\right)^{-N}\right\}
$$
from whence in a limit $N\to\infty$ one obviously obtains~(\ref{27}).

Hence all the path integrations have been made and we obtain for
the transition amplitude
\begin{eqnarray}
Z_\Psi=-T_\tau{}^2\int\frac{d^4 p}{(2\pi)^4}
e^{ip(x_f-x_i)}\frac{d\la_T d\la_S}
{(2\pi)^2}&\exp\left\{-i\frac{T}{2}\la_T(p^2+m^2)+i\la_S T_\tau
J\right\}\cdot\nonumber\\
&\cdot\exp\left\{2\ve_\z\z_1\hat p\bar\z_2
e^{-i\ve_\z\la_S T_\tau /2}\right\}\, . \label{28}
\end{eqnarray}
Now only integrations over zero modes remain to be performed.

To characterize the gauge group orbits in the extended phase space, we
introduce Teichm\"{u}ller parameters
\begin{equation}\label{29}
C_T=\frac{1}{2}\int\limits_{\tau_i}^{\tau_f}\la_T (\tau){}d\tau,\qquad
C_S=\frac{1}{2}\int\limits_{\tau_i}^{\tau_f}\la_S (\tau){}d\tau\, .
\end{equation}
The parameter $C_T$ has a transparent physical sense. In a suitable gauge it
is the proper time~\cite{9}. The parameter $C_S$ appears
due to the fact that internal quantum numbers,
such as a spin, a charge, etc. are realized in
classical terms as topological toroidal--path characteristics. Let the
parameter $C_S$ be called the proper spin phase angle. As a result of
boundary conditions on the parameters of reparametrization symmetry
$\e(\tau_i)=\e(\tau_f)=0$ and the phase transformations of index spinors
$\varphi(\tau_i)=\varphi(\tau_f)=0$, the Teichm\"{u}ller parameters cannot be
altered by gauge transformations because $\delta\la_T=\dot\e$,
$\delta\la_S=\dot\varphi$. Admissibility of using the gauge with derivatives
$\dot\la_T=\dot\la_S=0$ means that the gauge group orbits are bijectively
characterized by zero modes of the Lagrange multipliers, for which, obviously,
one has
\begin{equation}\label{30}
C_T=\la_T\cdot T_\tau/2,\qquad C_S=\la_S\cdot T_\tau/2.
\end{equation}

Since the evolution parameter must bijectively correspond to the points of
the particle world line~\cite{9}, only reparametrizations described by strictly monotonic
functions are admissible. As a consequence, the reparametrization group
falls into two connected components. One of them is the subgroup which
preserves the world line orientation, the second one is the set of
reparametrizations which change this orientation. The corresponding
modular group (the quotient of the
complete gauge group by the connected component of the
unit) is $Z_2$. The BFV--BRST quantization includes only gauge transformations
which are continuously connected with the identical one, so the integration is to be
taken over the fundamental domain of the modular group in the Teichm\"{u}ller
space. Let us choose the domain for the parameter $C_T$ assuming $C_T>0$,
then positive--energy particles move forward in time and the transition
amplitude~(\ref{8}) is the causal propagator.

If it is assumed that internal quantum numbers are independent of the state of
particle motion, then the fundamental domain of the modular group for phase
transformations of index spinors is obvious from the expression derived for
the amplitude~(\ref{28}). Owing to the integrand periodicity in the parameter
$C_S=\la_S T_\tau/2$ at half--integers $J$, any interval period in length,
say $[0,2\pi]$, can be taken as a fundamental domain. The modular group of
phase transformations is the group $Z$. One can invert the consideration
and regard the modular invariance of the transition amplitude as a condition
on the quantum theory obtained from the classical formulation by means of
path--integral calculation. Then the boundary conditions on the parameter
$\varphi$ should be weakened as  $\varphi(\tau_f)-\varphi(\tau_i)=2\pi n$,
$n\in Z$, and the requirement of single--valuedness for the transition
amplitude leads immediately to quantization of the spin $J$ (see
a consideration of similar type, e. g., in~\cite{25}).

In~(\ref{29}) integration over the Teichm\"{u}ller parameter $C_T$  is
performed by using the well--known equality
\begin{equation}\label{31}
\int_0^\infty\,dC_T\exp\{-iC_T(p^2+m^2)\}=-i/(p^2+m^2-i0).
\end{equation}
So, the choice of a fundamental domain is equivalent to the usual pole bypass
rule in the integral representation of the causal propagator.

The integral over the parameter $C_S$ is found by application of the Cauchy
integral formula $f^{(n)}(z)=\frac{n!}{2\pi i}\oint\limits_{{\cal C}}
\frac{f(z^\prime)}{(z^\prime-z)^{n+1}}dz^\prime$
for the $n$--th derivative of an analytic function $f(z)$
of a complex variable $z$ in the interior of the domain
bounded by a contour ${\cal C}$. If $f(z)=\exp (Az)$ and
the contour ${\cal C}$ is the unit circle with the center at
$z$, so that $z^\prime=z+e^{i\alpha}$ can be used as its parametrization,
then we easily arrive at
\begin{equation}\label{32}
\int_0^{2\pi}\,\exp(-in\alpha+Ae^{i\alpha})\,d\alpha=2\pi A^n/n!\, .
\end{equation}

Finally, integrating in~(\ref{29}) over Teichm\"{u}ller parameters with
the help of the found equalities~(\ref{31}) and~(\ref{32}), we obtain the
transition amplitude
\begin{equation}\label{33}
Z_\Psi=\frac{-i}{(2J\ve_\z)!}\int\,\frac{d^4 p}{(2\pi)^4}e^{ip(x_f-x_i)}
\frac{(2\ve_\z\z_1\hat p\bar\z_2)^{2J\ve_\z}}{p^2+m^2-i0}\cdot\frac{2}{\pi}\, ,
\end{equation}
which is nothing but the index--free form of Weinberg propagator~\cite{4}
received in the $(2J+1)$--component formalism of the field theory.
In the holomorphic case, the correct values of $J$ are positive~\cite{3},
$J\geq 0$, and particles are
described by symmetric spinors of rank $2J+1$ with undotted indices.
In the antiholomorphic case $J\leq 0$, and
particles are described by spinors with dotted indices.
In line with common reasons~\cite{9}, connection among the sign of $J$
and the sign of energy shows that alternation of the choice of boundary
conditions~(\ref{13}) is equivalent to alternation of the definition of particles
and antiparticles.

It should be noted that the spin dependent multiplier in
integrand~(\ref{33}) can be represented in the form
$$
\frac{(2\ve_\z\z_1\hat p\bar\z_2)^{2J\ve_\z}}{(2J\ve_\z)!}=
\frac{(2\ve_\z\z_1\hat p\bar\z_2)^{2J\ve_\z}}{\Gamma(2J\ve_\z+1)}
$$
which is unified for the whole spin tower. It indicates a possibility of analytic
continuation to ``any'' complex $J$~\cite{13,14}, this being important for the theory
of moving Regge poles and the string theory.

\section{The transition amplitude as an index--free form of the propagator}
Comparison of the obtained result with the result of paper~\cite{4} can be
realized as follows. For the sake of definiteness, we shall
restrict ourselves to the holomorphic case.
Characteristics of the Wigner wave function $u(p,\z;\sigma)$ are determined by
the primary quantization procedure~\cite{3}, thus it obeys the spin constraint
 $(\hat S_\z -J)=0$ and the spinor constraint $\hat{\bar d}_\z u=0$,
where the index spinor operators are realized as multiplication operators,
$\hat\z=\z$, and operators of their canonically conjugate momenta are
realized as differentiation operators, $\hat p_\z=-i\partial /\partial\z$.
As a consequence~\cite{3}, $u(p,\z;\sigma)=e^{-\z\hat p\bar\z}[\z]^{J,\sigma}$,
where $[\z]^{J,\z}$ is the homogeneous polinomial in $\z$ with degree $2J$,
$\sigma=-J,-J+1,\cdots,J-1,J$.

It is important that the Wigner wave function of arbitrary momentum can be
obtained from the wave function of the standard momentum by some
transformation of the index spinor only
$$
u(p,\z;\sigma)=u(\stackrel{o}{p},\z B_p;\sigma)\, .
$$
Here $B_p=B_p{}^+$ is the Wigner operator,
$\hat p=B_pB_p{}^+$, and $\stackrel{o}{p}=(m,0)$ is the
standard momentum. This circumstance makes possible such easy to pass from
the arbitrary momentum frame to the standard momentum frame and conversely that
in the following we usually do not thoroughly specify in what namely frame
the consideration is carried out.

In the rest frame, the polynomial $[\z]^{J,\sigma}$ satisfies just the same
condition as the Wigner wave function does, i. e.,
$(\hat M_3-\sigma)u(\stackrel{o}{p},\z;\sigma)=0$, where the spinor part
of the third component of the angular momentum is
$\hat M_3=\frac{1}{2}(-\z^1\frac{\partial}{\partial\z^1}+
\z^2\frac{\partial}{\partial\z^2}+c.c.)$. This equation determines the degree
$(J\mp \sigma)$ of the coordinate $\z^{1,2}$ entering into $[\z]^{J,\sigma}$,
hence $[\z]^{J,\sigma}=
N_J {{2J}\choose {J-\sigma}}^{1/2}(\z^1)^{J-\sigma}(\z^2)^{J+\sigma}$.
Here ${{2J}\choose {J-\sigma}}$ is a binomial coefficient which allows us to
identify the contraction over spinor indices and over the spin projection
$\sigma$. The normalization multiplier $N_J$ is found below.

For transition to an arbitrary frame of reference one has to use the
relation
$$
[\z B]^{J,\sigma}=[\z]^{J,\sigma^\prime}D^J(B)_{\sigma^\prime}{}^\sigma \, ,
$$
where $B$ is an arbitrary $2\times 2$ matrix and $D^J$ is the Wigner
D--function.

The standard sesquilinear form in the space of holomorphic functions of the index
spinor induces the inner product for polynomials in $\z$:
\begin{equation}\label{34}
(\varphi,\psi)=N\int{}d^2\z\,d^2\bar\z\,e^{-2\z\hat p\bar\z}\bar\varphi\psi.
\end{equation}
For homogeneous functions of degree $2J$ this inner product can be written in
terms of the differential operator
\begin{equation}\label{35}
(\varphi,\psi)=\frac{2^{-(2J+2)}}{(2J)!m^{4J}}(\frac{\partial}{\partial\z}
\hat p\frac{\partial}{\partial\bar\z})^{2J}\bar{\varphi^J}\psi^J\cdot
N\cdot\frac{4\pi^2}{m^2}\, .
\end{equation}
The right--hand side of~(\ref{35}) coincides (up to the multiplier)
with the known expression, see, e. g.~\cite{14} where the common factor is not fixed. Now
from the orthonormality condition
$([\z]^{J^\prime,\sigma^\prime},[\z]^{J,\sigma})=
\de_{J^\prime J}\de_{\sigma^\prime\sigma}$  the normalization
of basic symmetric spinors $[\z]^{J,\sigma}$ can easily be found.
It is sufficient
to restrict ourselves to the calculation for the values $\sigma=-J$:
\begin{equation}\label{36}
N_J{}^2=\frac{2^{2J+2}}{(2J)!m^{2J}}\cdot\frac{m^2}{4\pi^2}\cdot\frac{1}{N}.
\end{equation}
The normalization multiplier $N$ is found from the condition that the
expression~(\ref{36}) has to be equal to unity for $J=0$: $N=m^2/\pi^2$.

Then to obtain the Weinberg propagator it is necessary to integrate
the integrand~(\ref{33}) multiplied by
$[\z_i]^{J,\sigma^\prime}[\bar\z_f]^{J,\sigma}$ over initial
$\z_i$ and final $\z_f$ index spinors with the measure defined by Eq.~(\ref{34}).
In such a way we obtain the propagator~\cite{4}
\begin{equation}\label{37}
G_{\sigma^\prime\sigma}^J(x)=
-im^{-2J}\Pi_{\sigma^\prime\sigma}^{J}(i\partial)\Delta^C(x)\, ,
\end{equation}
where
$$
\Delta^C(x)=(2\pi)^{-4}\int{}d^4 p{}e^{ipx}/(p^2+m^2-i0)
$$
is the causal Green's function of a scalar field, and the $(2J+1)\times(2J+1)$~--~component
matrix $\Pi_{\sigma^\prime\sigma}^J$ is determined by identities in the following
chain
\begin{equation}\label{38}
\frac{1}{(2J)!}(2\z_f\hat p\bar\z_i)^{2J}=
\frac{1}{(2J)!}((\z_f B_p)2\hat{\stackrel{o}{p}}(B_p\bar\z_i))^{2J}
\equiv [\z_f B_p]^{J,\sigma^\prime}\Pi_{\sigma^\prime\sigma}^J
({\stackrel{o}{p}})[B_p\bar\z_i]^{J,\sigma}=
\end{equation}
$$
=[\z_f]^{J,\sigma^\prime}\Pi_{\sigma^\prime\sigma}^J(p)
[\bar\z_i]^{J,\sigma}\equiv
p_{\mu_1}\cdots p_{\mu_{2J}}[\z_f]^{J,\sigma^\prime}
t_{\sigma^\prime\sigma}^{\mu_1\cdots\mu_{2J}}[\bar\z_i]^{J,\sigma}(-1)^{2J}\, .
$$
The properties of $t_{\sigma^\prime\sigma}^{\mu_1\cdots\mu_{2J}}$ and
$\Pi_{\sigma^\prime\sigma}^J(p)$ have been described in detail in~\cite{4}.

In particular, it is essential in the calculation~(\ref{38}) that
the quantities $t_{\sigma^\prime\sigma}^{\mu_1\cdots\mu_{2J}}$ and
$\Pi^J_{\sigma^\prime\sigma}(p)$ have the following properties.\\
i) $t_{\sigma^\prime\sigma}^{\mu_1\cdots\mu_{2J}}$ is symmetric with respect to
the 4--vector indices, because it is defined by contraction with
the tensor power of the energy--momentum vector.\\
ii) $t_{\sigma^\prime\sigma}^{\mu_1\cdots\mu_{2J}}$ is traceless with respect
to the 4--vector indices, due to the identity
$\sigma_{\alpha\dot\alpha}^\mu\sigma_{\mu\beta\dot\beta}
=-2\e_{\alpha\beta}\e_{\dot\alpha\dot\beta}$
and automatic symmetrization of the tensor power of the commuting spinor
in spinor indices.\\
iii) $\Pi_{\sigma^\prime\sigma}^J(p)$ is a tensor, i. e.
$$
D^J(A)\Pi^J(p)D^J(A)^+=\Pi^J(p^\prime)\, ,
$$
where $A\in SL(2,C)$, $\hat p^\prime=A\hat p A^+$. The irreducibility of
the representation of the small group $SO(3)$, which follows naturally from
the model considered, and the Schur lemma mean that
$\Pi(\stackrel{o}{p})$ is a multiple
of the identity matrix and it is normalized as
$\Pi_{\sigma^\prime\sigma}^J(p)=m^{2J}\de_{\sigma^\prime\sigma}$.
It can be shown,~\cite{4}
that $\Pi(p)$ is a polynomial of degree $2J$ in the helicity operator
$\vec p\cdot\overrightarrow M^{(J)}/|\vec p|$. On the mass shell we have
$$
\Pi^J(p)=m^{2J}D^J(B_p)^2=
m^{2J}\exp\left(-2\theta\vec p\cdot\overrightarrow M^{(J)}/|\vec p|\right)\, ,
$$
where $\theta$ is defined by $\sinh\theta=|\vec p|/m$.
An explicit expression for the matrix $\Pi^J$
is given in~\cite{4}.
In the derivation~(\ref{37}) from~(\ref{33}), one should include the additional
multipliers $1/\pi$ (given by comparison with the direct calculation for
$J=0$) and $2i$ (found from comparison between expressions for $J\ne 0$),
which display the differences in the insertion of the pole multiplier
in the integrand and in the
transition to the nonzero spin case in our approach and in ref.~\cite{4}.

Now, the relation between expressions~(\ref{33}) and~(\ref{37}) is obvious.
\section{Conclusion}
Thus, as it should be expected, the above--obtained transition
amplitude~(\ref{33}) coincides
with the index--free form of the Weinberg
propagator~(\ref{37})~\cite{4} for the massive
particle with any spin $J$, found
in the ($2J+1$)--component formalism of the field
theory. This result is obtained  with the use of the BFV--BRST
path--integral approach for the first time. It should be noted that it has
been obtained without arbitrary renormalizations of the path integral
measure. A similar study of the massless spinning particle,
the spinning particle
in the formulation with Dirac index Bose--spinors (the $2(2J+1)$--component
formalism of the field theory) and for the higher space--time dimensions,
as well as the supersymmetric generalization will be the subject of further
articles.
\vspace{0.12cm}

The authors are grateful to I.A.Bandos, D.P.Sorokin, A.Yu.Nurmagambetov for the
hospitality at the NSC Kharkov Institute of Physics and Technology,
Kharkov, Ukraine, and for
fruitful discussion. We would like to thank E.A.Ivanov and S.O.Krivonos
for the hospitality at
the Joint Institute for Nuclear Research, Dubna, Russia,
and A.I.Pashnev as well for providing us with an important reference.
One of us (V.Z.) thanks A.A.Kapustnikov for the productive discussion
on $\kappa$~--~symmetry.

\end{document}